\documentstyle[prl,aps,psfig]{revtex}

\newcommand{\be}{\begin{equation}}
\newcommand{\ee}{\end{equation}}
\newcommand{\fig}[1]{fig.~\ref{#1}}

\begin{document}

\draft

\twocolumn[\hsize\textwidth\columnwidth\hsize\csname@twocolumnfalse\endcsname

\title{Growth of Patterned Surfaces}
\author{Harald Kallabis and Dietrich E.~Wolf}
\address{H\"ochstleistungsrechenzentrum, Forschungszentrum J\"ulich,
D-52425 J\"ulich, Germany\\and\\
Theoretische Physik, FB10, Gerhard--Mercator--Universit\"at Duisburg,
D-47048 Duisburg, Germany}

\maketitle

\begin{center}
September 5, 1997
\end{center}

\begin{abstract}

During epitaxial crystal growth a pattern that has initially been
imprinted on a surface approximately reproduces itself after the
deposition of an integer number of monolayers. Computer simulations of
the one-dimensional case show that the quality of reproduction decays
exponentially with a characteristic time which is linear in the
activation energy of surface diffusion.  We argue that this life time
of a pattern is optimized, if the characteristic feature size of the
pattern is larger than $(D/F)^{1/(d+2)}$, where $D$ is the surface
diffusion constant, $F$ the deposition rate and $d$ the surface
dimension.

\end{abstract}

]

\narrowtext

Modern techniques of manipulating crystal surfaces allow to imprint
structures down to the atomic size on them. With the tip of a
tunneling microscope one can arrange adsorbed atoms in a pattern. In
heteroepitaxial growth a two dimensional array of quantum dots may
form on the nanometer scale. With masking techniques arbitrary
patterns with features as small as a micrometer along the surface and
atomic size perpendicular to it can be fabricated. If one buries such
a pattern under an overlayer growing in layer-by-layer mode, the
pattern will approximately reproduce itself periodically at the
completion of each layer. The question arises, how this propagation of
a pattern is influenced by the growth conditions. A theoretical
understanding of such temporal correlations began to emerge only
recently \cite{Somfai,Damping,Pers,Krech,Tersoff}. In the following
the propagation of a pattern will be discussed for the simplest case
that the influence of elastic strain on surface diffusion is
negligible and that diffusion across a step edge from an upper to a
lower terrace is not inhibited by Ehrlich-Schwoebel barriers \cite{ESB}.

Experimentally, it has long been known that layer-by-layer (or
Frank-van der Merwe) growth persists up to a time $\tilde t$, after
which the characteristic growth oscillations are damped out and the
surface becomes rough.  Recently, it has been shown that $\tilde t$
depends on the growth conditions \cite{Damping,Brendel,Wolf,Seoul},
i.e. the surface diffusion constant $D$ and the deposition rate $F$,
with a power law $(D/F)^{\delta}$.
Clearly, a pattern can at most survive as long as the surface grows
layerwise. In fact the life time of a pattern is much shorter than
$\tilde t$, as we are going to show: It depends on $D/F$ only
logarithmically rather than with a power law.

The life time of a pattern in layer-by-layer growth depends also on
the feature size $r$. We shall identify the length scale that is
important in the context of pattern decay as
$\ell_0\sim(D/F)^{1/(d+2)}$. For feature size $r \gtrsim \ell_0$, the
life time increases rather abruptly.

{\em Propagation probability.}
To simplify the discussion, we assume that the grown crystal is simple
cubic \cite{SquareIn2D} and that overhangs and defects can be
neglected.  Then, the surface configuration at a given time $t$ can be
represented by a height function $h(x,t)$ with $x$ indexing the
lattice sites on a $d$--dimensional substrate. In the following, time
will be measured in numbers of deposited monolayers.

When the growth process is initiated at time $t=0$ with some structure
$h(x,0)$, consisting of e.g.\ islands of a particular shape on an
otherwise flat substrate (a {\em pattern~}), new islands will form on
top of the initial ones, and all of them will expand laterally due to
the attachment of atoms at their edges. Therefore the pattern will be
deformed during this early stage of growth.
However,  the original pattern will nearly be reproduced after the
deposition of one monolayer, because the new layer nucleates
preferentially near the centers of islands in the previous layer. As
this correlation extends over long times \cite{Somfai}, one expects
the approximate reproduction of the pattern at later times as well. 
It makes sense to ask, which fraction of the pattern is propagated
through $t$ monolayers. We call this fraction the propagation
probability $p(t)$. It is given by
\be
\label{decaydef}
p(t) \equiv \langle \prod\limits^{t}_{s=1} 
\delta_{h(x,s),h(x,0)+s} \rangle  .
\ee
The brackets denote averaging over different lattice sites $x$ and
different realisations of the growth process.
$\delta_{i,j}=1$ if $i=j$ and 0 otherwise denotes the Kronecker delta.

By defining $p(t)$ in this way, we measure the {\em deterministic~}
reproduction of the initial pattern $h(x,0)$, seen stroboscopically
after deposition of $1, 2, \dots$ monolayers in the comoving frame.
It is {\em deterministic} in the sense that a given
site $x$ is counted as 'surviving' after time $t$ only if it has
survived through {\em all~} previous times $1,\dots,t$. Of course, the
height might also regain its initial value by chance once it has left
it, but this is a stochastic process and contributes to
the noisiness of the pattern. Thus the survival of {\em information} is not
described properly by the propagation probability (\ref{decaydef}),
but by an appropriate entropy. It will be dealt with in a longer paper.

If the probability of return to the initial height (in the comoving
frame) at
time $t$ was considered, irrespective of the values at intermediate
times, one would probe a two-point function. Such a 
two--point function is expected to decay algebraically for long
times. More precisely, it should scale like the value of the height
distribution function at the average height. For self-affine surfaces
the width of the height distribution increases like $t^\beta$. Thus
the probability to recover the initial height 
after the deposition of $t$ monolayers, should decay as
$t^{-\beta}$.

\begin{figure}[htb]
\centerline{\psfig{figure=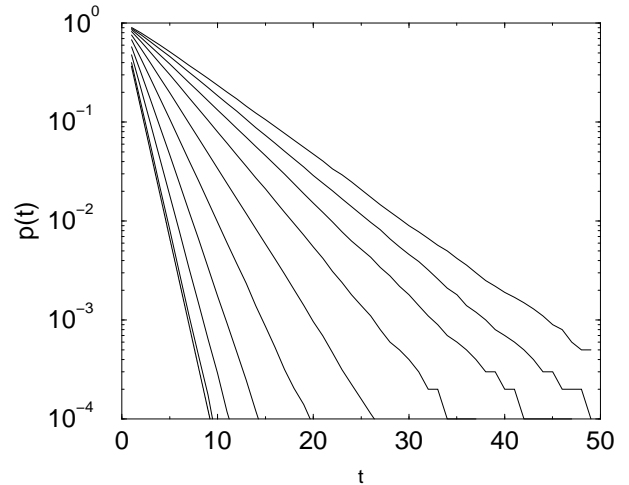,width=8.0cm,angle=270}}
\caption{
  Propagation probability $p(t)$ of a flat surface as function of
  number of deposited monolayers $t$ for different values of $D/F=0,
  10^0, 10^1, \dots, 10^8$ from left to right.  The characteristic
  time of the exponential decay increases with increasing $D/F$. 
}
\label{expdecay.fig}
\end{figure}

In contrast to this, the $t-$point function $p(t)$
decays exponentially, see \fig{expdecay.fig}. Although the
exact evaluation is non-trivial, this result is easily made plausible:
The fraction $p(t+1)$ of surviving sites at time $t+1$ equals
the number of surviving sites at time $t$, $p(t)$, times the
probability to propagate to the next layer. Assuming that this
probability can be identified with $p(1)$, independent 
of the surface configuration, which at time $t$ of course differs from
the initial one, the exponential decay
\be
\label{expdecay}
p(t)=\exp(-t/t_c),
\ee
follows immediately. We shall show below that the propagation
probability does depend on the surface configuration. However,
this dependence is so weak, that the surface evolution during
the life time $t_c$ of a pattern hardly affects its exponential decay
(\ref{expdecay}).

The main purpose of this Letter is to investigate the dependence
of $t_c$ on the microscopic growth parameter $D/F$ and the 
feature size of the pattern. Let us first discuss two limit cases.
For $D/F\to 0$, the sites are not coupled by diffusion and
the appropriate description of the growth process is given by the 
random deposition model \cite{RD}, where atoms are deposited randomly
onto the substrate and remain at the deposition site forever.  In this
model, the fraction $S_j(t)$ of surface sites in layer $j$ after time
$t$ is a Poisson distribution, $S_j(t) = \exp(-t) t^j / j!$.
Therefore,  $p(1) = S_1(1) =
\exp(-1)$, so that $t_c=1$.

For $D/F\gg1$, a monotonous increase of $t_c$ as function of $D/F$ is
expected. When the initial 'pattern' is simply a flat surface, one
expects $t_c\to\infty$ for $D/F\to\infty$, because layer-by-layer
growth persists forever for infinitely high diffusion constant. The
computer simulations of molecular beam epitaxy (MBE) on a
one--dimensional substrate, to 
be presented in the next section, confirm this picture and show a
dependence
\be
\label{logdep}
t_c \sim \log(D/F)
\ee
with a cutoff at $t_c=1$ for small $D/F$.

In the following we shall discuss three different initial
patterns: (1) A completely flat surface, (2) a rough surface as it
evolves from the flat one after time $\tilde t$, when the oscillations
due to layer-by-layer growth have died out,
and (3) a periodic modulation of the surface with a fixed
feature size. The first arises as a natural limit of a pattern with a
characteristic feature size $r\to\infty$. 
The second represents the simplest generic configuration for which the
growth kinetics has no periodic time dependence any more.
Both will be used to study the pattern decay process systematically in
the next section.
Finally, the third choice will lead to the optimization condition for
pattern survival and will be studied afterwards.

{\em Model and simulation results.} 
Atoms are deposited onto a one--dimensional substrate of typical size
$L=10^4$ with a rate of $F$ atoms per unit time and area.  Atoms with
no lateral neighbour are allowed to diffuse with diffusion constant
$D$. Atoms with lateral neighbours are assumed to be immobile, so that
e.g.\ dimers are immobile and stable. 
Growth commences with a flat
substrate, $h(x,0)=0$ for all sites $x=1,\dots,L$. (The other initial
configurations will be discussed below.) On deposition at $x$,
$h(x,t)$ is increased by one. 

\begin{figure}[htb]
\centerline{\psfig{figure=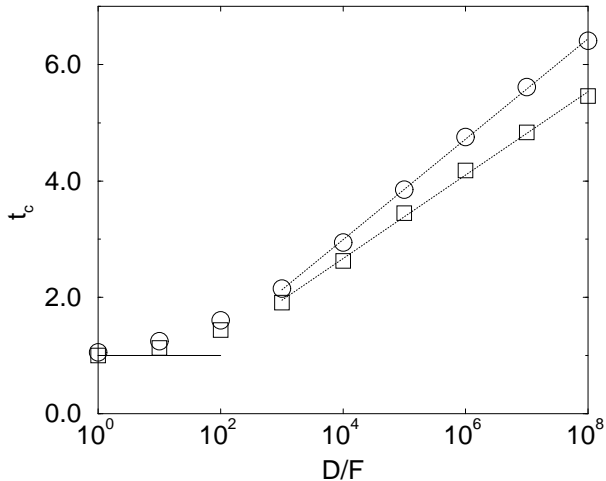,width=8.0cm,angle=270}}
\caption{
  Pattern life time $t_c$ as function of $D/F$ for a flat 
  ($\circ$) and a rough ($\Box$) surface.
  The solid line indicates the
  minimum life time $t_c=1$ for $D/F\to0$. The dotted
  lines represent logarithmic fits to the last five decades of $D/F$.
}
\label{FtcDdF.fig}
\end{figure}

Fig.~\ref{expdecay.fig} confirms the exponential decay (\ref{expdecay})
of the propagation probability.
The initial configuration survives the better the higher the value of
$D/F$, as expected. The life times $t_c(D/F)$ are shown in
\fig{FtcDdF.fig}. For small $D/F$ the life time approaches 1
as derived above for random deposition, and for large $D/F$ it increases
like $\log(D/F)$. 

The results for a rough surface as initial pattern are very similar.
The propagation probability (\ref{decaydef}) was averaged over 100
different initial patterns, all obtained by depositing 50,
respectively 200  monolayers on a flat substrate. For the values of
$D/F$ considered here the periodic oscillations of the surface
morphology have stopped by then, as shown in
\cite{Damping}. Interestingly, the life time of a rough surface pattern
is shorter than that of the flat surface, see \fig{FtcDdF.fig},
but we could not observe any difference between the rougher surface
(200 monolayers) and the less rough one (50 monolayers). This will
become plausible, when considering, how the life time of a pattern is
affected by the feature size.

{\em Periodic patterns.}
In this section, we study the dependence of the life time on the
typical feature size of an artificially prepared initial configuration.
To this end, it is useful to recall the different length scales
associated with ideal MBE:

The island distance or diffusion length $\ell$ is a function of $D/F$:
\be
\ell \sim (D/F)^\gamma.
\ee
The exponent $\gamma$ depends on the substrate dimension, on the size
of the critical nucleus and the possible fractality of the islands
\cite{Wolf,Seoul,Gamma}. Its numerical value for the simulations
presented here is $\gamma=1/4$.

The only dimensionless length scale $\ell_0$ which can be
constructed from the dimensionful parameters $D$ and $F$, is
\be
\label{l}
\ell_0 \sim (D/F)^{1/(d+2)}.
\ee
Physically, this length scale comes from comparing the diffusion time
to the adatom arrival time on an area of size $l^d$
\cite{Wolf,Schroeder}. $\ell_0$ and $\ell$ are submonolayer
quantities.

Finally, the layer coherence length $\tilde \ell$ depends on $D/F$ like
\be
\label{ltilde}
\tilde \ell \sim (D/F) ^{4\gamma/(4-d)},
\ee
see \cite{Damping}. This is not a submonolayer quantity, as $\tilde
\ell$ appears as the typical length scale after the oscillation damping
time, i.e.\ after deposition of $\tilde t$ monolayers.

In order to study the length scale dependence of the life time, 
we use a periodically modulated surface
\be
h(x,t=0) = \Theta\left(\sin(\pi x/r)\right)
\ee
where $\Theta(x)=1$ if $x\ge0$ and 0 otherwise is the Heavyside function.

\begin{figure}[htb]
\centerline{\psfig{figure=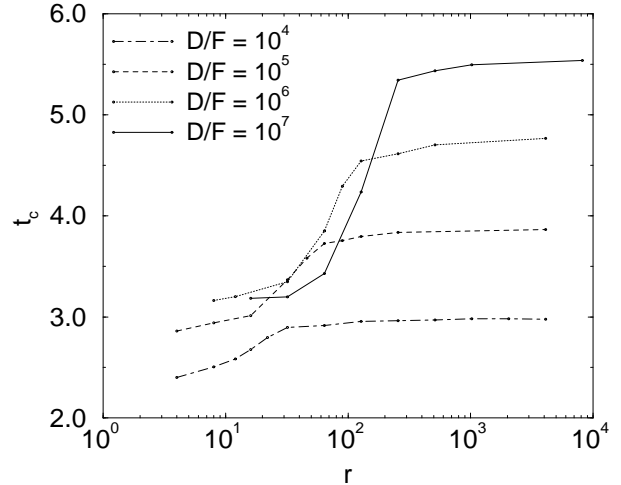,width=8.0cm,angle=270}}
\caption{
  Life time $t_c$ as function of the feature size $r$ of the initial
  surface modulation for $D/F=10^4, 10^5, 10^6, 10^7$. Around a
  characteristic feature size depending on $D/F$, the life time
  increases rapidly.
}
\label{inilen.fig}
\end{figure}

The measurements of the life time as a function of the initial
wavelength $r$ for different $D/F$, see \fig{inilen.fig}, show that
for $r$ greater than a characteristic value depending on $D/F$,
survival is strongly enhanced. Above this value, the life time depends
only little on the feature size. For small $r$, the life time seems to
saturate for increasing $D/F$. These findings can be understood in the
following way.

The mechanism of transporting the memory of the surface structure from
one monolayer to the next, is that nucleations take place near the
center of islands that have already formed one layer below. If the
feature size $r$ is chosen so small that nucleations cannot take place
on top of the pattern, this mechanism is suppressed and consequently
the life time of the pattern is reduced. To suppress nucleations, the
distance between sinks for adatoms, i.e.\ the feature size $r$, has to
be chosen so small that a freshly deposited adatom diffuses to the
nearest sink (i.e. a distance $r$), before the next atom is deposited
within the area $\sim r^d$. This is the case for $r \lesssim \ell_0$.
Hence, $\ell_0$ should be the length scale found in \fig{inilen.fig}.

\begin{figure}[htb]
\centerline{\psfig{figure=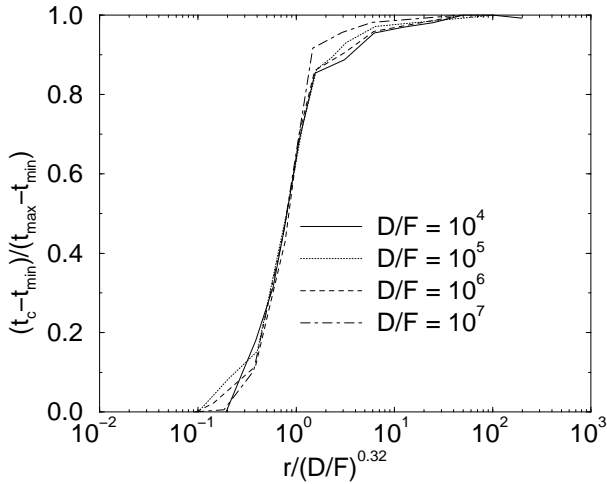,width=8cm,angle=270}}
\caption{
  Data from \fig{inilen.fig} with normalized life times, and feature
  size $r$ scaled with $(D/F)^{0.32}$.
}
\label{inilen_scal.fig}
\end{figure}

The scaling plot \fig{inilen_scal.fig} shows that the characteristic
length above which the survival is prolonged, scales like
$(D/F)^{0.32\pm0.01}$, which suggests a length scale proportional to
$(D/F)^{1/3}$. This is in accordance with the argument given above for
the characteristic length scale being $\ell_0$.  To assure that the
characteristic length scale in \fig{inilen.fig} is not $\tilde \ell$,
which for the parameters studied here also is proportional to
$(D/F)^{1/3}$, we studied epitaxial growth with a critical nucleus of
2 instead of 1, which influences $\tilde \ell$, but not $\ell_0$. This
analysis shows that indeed $\ell_0$ is the characteristic length
scale.  Details on this will be published in a longer paper.

The saturation of $t_c$ for small $r$ as function of $D/F$ can now
easily be understood: If $r\ll\ell_0$, the memory of the initial
periodic pattern is destroyed already after the deposition of one
monolayer. The surface will be half filled with islands which
have a typical size $\ell$ much larger than $r$. Then the fraction
of sites which propagated by just one lattice constant is about 50\%,
irrespective, how large $D/F$ is. 

The faster decay of a rough compared to a flat surface as initial pattern
can be made plausible with the following reasoning:
The feature size of rough surfaces may be identified with the typical
terrace size $\ell$. For the simulations presented here, where the
critical nucleus was 1, $\ell$ is smaller than $\ell_0$. Therefore the
faster decay of the pattern is consistent with our findings for
periodic patterns.

{\em Conclusions and outlook.}
In conclusion we have shown that a pattern decays exponentially fast
with a life time proportional to $\log(D/F)$. With the Arrhenius law,
$D\sim\exp(-E/k_BT)$, the life time 
decreases linearly with the energy barrier $E$ for surface diffusion. The
life time of a pattern is optimal if the feature size of the pattern
is larger than $(D/F)^{1/(d+2)}$.

An important extension of this work would be the study of the
two--dimensional case. Whereas it is natural to expect an exponential
decay of the propagation probability, the dependence of the life
time on the microscopic growth parameters is an open question. 

In this paper, we neglected barriers for interlayer transport
(Ehrlich--Schwoebel barriers) \cite{ESB}. The memory mechanism
will be enhanced by them, but the instability \cite{villain91}
associated with them 
will tend to make pattern reproduction worse. The competition between
these two effects is well worth studying as in many materials
interlayer transport in inhibited.

Useful conversations with J\'anos Kert\'esz, Joachim Krug and Martin
Rost are gratefully acknowleged. This work was supported by the
Deutsche Forschungsgemeinschaft (SFB 166).


\begin{references}
\bibitem{SquareIn2D}
  Our simulation data are obtained for the two--dimensional case of a
  square lattice.
\end{references}
\end{document}